\documentclass[10pt,journal,final,twocolumn]{IEEEtran}
\usepackage{multicol}
\usepackage{multirow}

\usepackage{amsmath}
\usepackage{subfigure}

\usepackage{setspace} %Ê¹ÓÃ¼ä¾àºê°ü
\usepackage{psfrag}
\usepackage{amsfonts}
\usepackage{amssymb}
\usepackage{stmaryrd}
\usepackage{mathrsfs}
\usepackage{epstopdf}
\usepackage{enumerate}
\usepackage{booktabs}  % ÈýÏß±í¸ñ
\usepackage{amsmath, color}
\usepackage{dsfont,bm}
\usepackage{amssymb}
\usepackage{epsfig}
\usepackage{graphicx}
\usepackage{amssymb}
\usepackage{epsfig}
\usepackage{epstopdf}
\usepackage{subfigure}
\usepackage{balance}
\usepackage{multicol}
\usepackage{enumerate}
\usepackage{verbatim,algorithm,algorithmic,cite}
\usepackage{booktabs}

\usepackage{multirow}
\usepackage[table,xcdraw]{xcolor}
\ifCLASSINFOpdf
\else
\fi
\makeatletter

\newcommand{\Rmnum}[1]{\expandafter\@slowromancap\romannumeral #1@}
\makeatother

\begin{document}
%\title{Spectrally Compatible Waveform Design for MIMO Radar Multi-Target Detection in the Presence of Signal-Dependent Interference}
\title{Hybrid Beamforming in mmWave Dual-Function Radar-Communication Systems: Models, Technologies, and Challenges}

\author{\IEEEauthorblockN{Ziyang Cheng,~\IEEEmembership{Member,~IEEE}, Linlong Wu,~\IEEEmembership{Member,~IEEE}, Bowen Wang,~\IEEEmembership{Graduate Student Member,~IEEE}, Bhavani Shankar,~\IEEEmembership{Senior Member,~IEEE}, Bin Liao,~\IEEEmembership{Senior Member,~IEEE}
and Bj\"{o}rn Ottersten,~\IEEEmembership{Fellow,~IEEE}\vspace{-2em}}\\
%\thanks{xxx}
\thanks{Z. Cheng  and B. Wang  are with School of Information \& Communication Engineering, University of Electronic Science and Technology of China, Chengdu 611731, China
(e-mail:  zycheng@uestc.edu.cn B\_W\_Wang@163.com).}
\thanks{L. Wu, B. Shankar and B. Ottersten are with the Interdisciplinary Centre for Security, Reliability and Trust (SnT), University of Luxembourg, Luxembourg (e-mail: \{linlong.wu, bhavani.shankar\}@uni.lu). 
%Their work is supported in part by ERC AGNOSTIC under grant EC/H2020/ERC2016ADG/742648 and in part by FNR CORE SPRINGER under grant C18/IS/12734677.
}
\thanks{B. Liao is with the Guangdong Key Laboratory of Intelligent Information
Processing, College of Electronics and Information Engineering, Shenzhen
University, Shenzhen 518060, China (e-mail: binliao@szu.edu.cn).}
%\thanks{This work was presented, in part, at the 2018 ICASSP \cite{ChengICASSP2018}, and in part at the 2018 DSP \cite{CHENG2018DSP}.}
}

\maketitle
%{Shell \MakeLowercase{\textit{et al.}}: Bare Demo of IEEEtran.cls for IEEE Transactions on Magnetics Journals}



% for Transactions on Magnetics papers, we must declare the abstract and
% index terms PRIOR to the title within the \IEEEtitleabstractindextext
% IEEEtran command as these need to go into the title area created by
%\IEEEtitleabstractindextext{%
\begin{abstract}
As a promising technology in beyond-5G (B5G) and 6G, dual-function radar-communication (DFRC) aims to ensure both radar sensing and communication on a single integrated platform with unified signaling schemes. To achieve accurate sensing and reliable communication, large-scale arrays are anticipated to be implemented in such systems, which brings out the prominent issues on hardware cost and power consumption. To address these issues, hybrid beamforming (HBF), beyond its successful deployment in communication-only systems, could be a promising approach in the emerging DFRC ones.
%Dual-function radar-communication (DFRC) has recently emerged as a candidate beyond 5G (B5G) and 6G technology, achieving radar sensing and communication via common signaling and platform.
%Based on the DFRC system, it is anticipated that the large-scale array will be adopted to achieve accurate sensing and reliable communication links. To address the cost and power issues brought by the large array, hybrid beamforming (HBF), beyond its successful deployment in communication-only systems, could be a promising approach in the emerging DFRC ones.
%To attain better localization performance and higher spectrum efficiency, the mmWave system with a large-scale array is envisioned as a potential scheme. 
%In mmWave systems, the hybrid beamforming (HBF) architecture, which includes large number of analog phase shifters (PSs) and small number of radio frequency (RF) chains, is a promising solution to reduce the cost and power consumption.   
In this article, we investigate the development of the HBF techniques on the DFRC system in a self-contained manner. % with the highlight of its key challenges and opportunities.
Specifically, we first introduce the basics of the HBF based DFRC system, where the system model and different receive modes are discussed with focus.
Then we illustrate the corresponding design principles, which span from the performance metrics and optimization formulations to the design approaches and our preliminary results. 
Finally, potential extension and key research opportunities, such as the combination with the reconfigurable intelligent surface, are discussed concisely.
% Dual-function radar-communication (DFRC),   which achieves the two  functionalities via using a common signaling and platform, has attracted considerable interest from both the academic and industry communities.  
% To attain better localization performance and higher spectrum efficiency, the mmWave system with a large-scale array is emerged as a potential scheme. 
% In mmWave  systems, the hybrid beamforming (HBF) architecture, which includes large number of analog phase shifters (PSs) and small number of radio frequency (RF)  chains, is a promising solution to reduce the cost and power consumption.   
% In this article, we introduce the HBF technology in mmWave DFRC systems, and discuss its  
% challenges and opportunities. 
% We first present the basics of HBF in DFRC systems. 
% Then we discuss the design principles of HBF in  DFRC systems in detail, including performance metrics, multi-objective optimization formulation,  design approaches and our preliminary results. 
% Finally, several challenges and frontiers in advanced HBF-based DRFC systems are pointed out. \textcolor{red}{the abstract should be improved}
\end{abstract}


\IEEEpeerreviewmaketitle

\IEEEpeerreviewmaketitle

\vspace{-1em}
\section{Introduction}
\IEEEPARstart{I}{n} the emerging beyond 5G (B5G) and sixth-generation (6G) networks, sensing capabilities have been recognized as a novel feature.
 % Sensing-as-a-Service (SaaS) has been recognized as a key component to enable various applications such as industrial Internet-of-Things (IoT), smart city and autonomous driving.
  The embedding of sensing functionality into wireless communication networks leads to the so-called dual-function radar-communication (DFRC), which has become  research area with significant activity of late \cite{LiuIntegrated2022}.
Unlike the solutions for co-existence of radar and communications, which mainly target spectrum sharing, DFRC integrates the radar and communication functionalities into a single platform under the umbrella of a unified signal processing framework. Not only can it improve the energy and spectrum efficiencies, but can also reduce hardware and signaling costs. In addition,  sensing capabilities can be used to enhance the performance of the wireless communication network itself by providing optimization input for network steering \cite{Zhang2022Enabling}.

%This aligns with the growing need for green network and addressing the spectrum congestion\cite{LiuJoint2020}.
%Profiting from sharing the same platform and framework, waveforms transmitted by DFRC with jointly designed, delicately optimized, bring satisfactory sensing and communication performance for various newborn technologies, which captures a booming interest from academia and industry in the DFRC systems.

With the expansion of spectrum allocations towards higher frequencies, such as millimeter-wave (mmWave) and terahertz (THz) bands, severe free-space path loss must be considered to build DFRC systems for  B5G and 6G. Massive multiple-input multiple-output (MIMO) with a large number of antennas is one of the key technologies to cope with this shortcoming. Nevertheless, if each antenna is equipped with a separate radio frequency (RF) chain to perform fully digital beamforming, the hardware cost and energy consumption could be prohibitive. Hence, hybrid beamforming (HBF), which shows great success in wireless communication \cite{Heath2016Overview}, has also been recognized as a promising trend for DFRC system development.

Essentially, the HBF technology uses a combination of low-dimension digital beamforming and high-dimension analogy bemaforming to attain a judicious trade-off between the system performance and hardware cost \cite{8030501}.  Specifically, only a small number of RF chains are deployed for digital beamforming, %and thus reduce the hardware cost and energy comsumption,
but the analogy beamforming is realized by many cost-efficient phase shifters (PSs).
Although HBF has been widely studied for communications, %in the context of communication-only systems,
it is still in its infancy for DFRC and further investigations are expected. Research questions like, how to model the HBF problem for DFRC and how the models impact on the principles and methodologies on system design are yet to be addressed. This dual-functional system requires new signal processing approaches. Additionally, practical constraints imposed on the DFRC system and coupling relationship between digital and analog beamformers will bring challenges to tackle high-dimension and non-convex optimization problems involved in various designs.
%, the design of HBF based DFRC system may result in a high dimension and non-convex optimization problem.
%
%
%the HBF in DFRC is still in its infancy and new challenges have arise in the investigation of the HBF based DFRC system.
%
%\textcolor{cyan}{
%Specifically, how the hybrid beamforming for DFRC is modeled crucially impacts the principles and methodologies applicable to DFRC systems design, which has been rarely considered in the existing literature.
%More importantly, since DFRC shares the signal processing framework, the existing processing approaches do not work well, requiring a new workflow to handle it.
%Moreover, due to practical constraints on the DFRC system and the coupling relationship between digital and analog beamformers, the design of HBF based DFRC system may result in a high dimension and non-convex optimization problem.
%}

While the extension of HBF from communication to DFRC is natural, it demands significant effort to pave its way for incorporation in future communication networks.
%, to ensure a successful deployment of the HBF structure on the emerging DFRC system, these challenges shall be addressed properly in the current and future research.
Building on the state-of-the-art  in this area, the afore-mentioned issues regarding on the models, technologies and challenges in HBF based DFRC systems will be discussed in this paper.  We hope this work can provide the reader an overview  of this important topic and encourage research towards further integration of sensing and communications in large antenna systems.

%{we illustrate various HBF-DFRC issues and organize them compactly in this paper, where the basics, design principles and new research frontiers will be introduced. We hope this work can provide you a bird's-eye view on this important topic in the ISAC trend.}

%The rest of this article is organized as follows. In Section II, we give the basics of HBF based DFRC system. Then we illustrate the corresponding design principles in Section III. Some simulation results are provided in Section IV. Some research frontier related to HBF are introduced in Section V, followed by the conclusions in Section VI.

% \begin{figure*}[t]
% 	\centering
% 	\includegraphics[width=0.85\linewidth]{./pic/OFDM_Diagram}
% 	%\vspace{-0.8em}
% 	\caption{Overview of a wideband OFDM-DFRC system  with hybrid analog and digital beamforming architectures at the transmitter and receiver.}
% 	%\vspace{-0.5em}
% 	\label{fig:pic1}
% \end{figure*}

%%%%%%%%%%%%%%%%%%%%%%%%%%%%%%%%%%%%%%%%
\section{Basics of Hybrid Beamforming in DFRC}
In this section, we introduce the basics of HBF in DFRC system, which will serve as the preliminaries for next sections. Specifically, we introduce the system model of HBF based DFRC system, followed by detailed DFRC receive modes.% will be illustrated concretely.

 \begin{figure}[t]
    \vspace{-1em}
     \centering
     \includegraphics[width=1\linewidth]{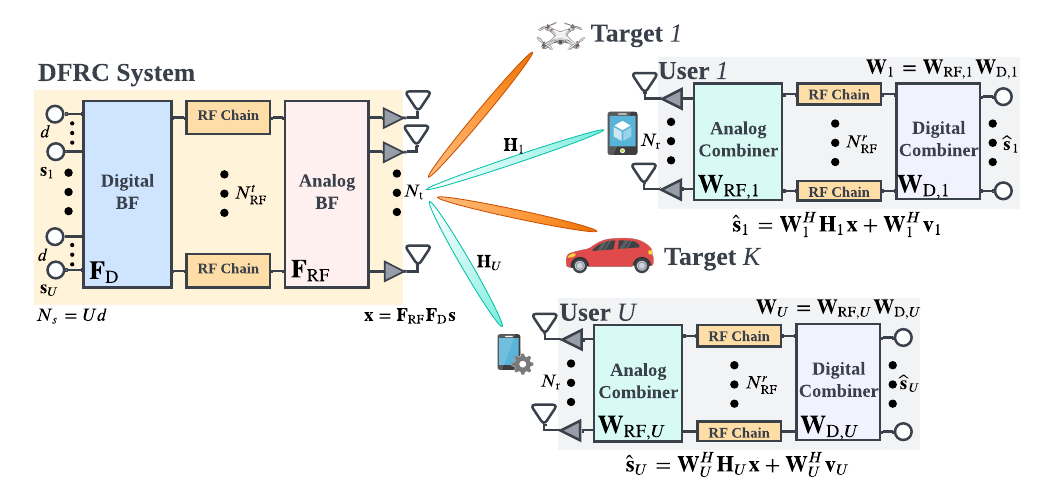}
     \caption{Overview of the hybrid beamforming based   DFRC system.}
     \label{fig:DFRC_multi_carrier}
 \end{figure}

\subsection{System Model of HBF based DFRC}
Fig. \ref{fig:DFRC_multi_carrier} gives the overview of the HBF-based DFRC system. There are $N_{\rm RF}^t$ RF chains and $N_t$ antennas at the transmitter. For the $U$ downlink users, each of them has $N_{\rm RF}^r$ RF chains and $N_r$ antennas.
For data transmission, the vector of all data symbols ${\bf s}$ has a length of $N_s$ and each symbol is usually assumed to be
statistically independent%, i.e.,  ${\mathbb E}\{ {\bf s}{\bf s}^H\}={\bf I}_{N_s}$
. The symbols to be transmitted are processed by a digital beamformer ${\bf F}_{\rm D}$, then up-converted to RF domain and precoded by an analog beamformer ${\bf F}_{\rm RF}$, which is realized by PSs. %and have a constant modulus $1/\sqrt{N_t}$. 
%Therefore, the emitted signal from the transmit antennas is $ {\bf x} =  {\bf F}_{\rm RF}  {\bf F}_{{\rm D}} {\bf s}$.

{For analog beamformers, there are three common architectures, i.e., full connection, partial connection and dynamic connection, as depicted in Fig. 2. More specifically, in the full connection architecture, each RF chain is connected to all antennas via PSs. For the partial connection structure, each RF chain connects to a fixed subset of antennas. As for the dynamic connection structure, each RF chain can dynamically connected a subset of antennas. A qualitative comparison of the analog beamformer structures is shown in Fig. 2(d), which indicates that by properly choosing the PSs network structure, the DFRC can achieve a good trade-off between performance and energy efficiency.
}

For communications, to estimate the symbols to user $u$, the received signal will be processed by both the analog and digital combiners as shown in Fig. 1. With the existence of noise on the receiver, {\color{black}the signal to user $u$ is processed by the analog combiner ${\bf W}_{{\rm RF},u}$
%${\bf W}_{{\rm RF},u}\in {\mathbb C}^{N_r \times N_{\rm RF}^r}$ 
followed by the digital combiner ${\bf W}_{{\bf D},u}$. 
Thus, by designing the HBF on both the transmit and receive sides jointly, a satisfying symbol error rate 
%(i.e., ${\mathbb E}\{ \|{\bf s}_u-\hat{\bf s}_u  \|^2  \}  $) 
can be ensured}\footnote{\color{black}The HBF design requires a  known  CSI, which is challenging to obtain for the HBF-based system due to the fact that the digital baseband has no direct access to the entries of the CSI matrix. Compressed
sensing-based methods can be used to achieve the channel estimation due to the sparse property 
of mmWave channels in the angle domain.}.

%  \cite{Yu2018JSTSP}
For radar sensing, the transmit beampattern should concentrate on the potential directions of the targets of interest while suppressing its sidelobes.
In the context of HBF, the transmit beampattern of the emitted signal is a function of the digital and analog beamforming matrices. %\cite{Cheng2021TCCN}.
Thereby, by designing the HBF properly, a 
desired transmit beampattern can be achieved.
%Upon the emitted signal $\bf x$, the transmit beampattern of the emitted signal, which is defined as the power of the transmit  signal at the direction $\theta$, can be formulated as a function of  ${\bf F}_{\rm D} $ and ${\bf F}_{\rm RF} $  \cite{Cheng2021TCCN}.  For radar sensing, the beampattern should concentrate on  the potential directions of the targets of interest while suppressing its sidelobes. Thereby, a desired transmit beampattern can be achieved by properly designing the HBF.
Additionally, for radar parameter estimation, if we consider a multi-carrier signaling scheme, we can obtain $K M_r$ virtual data vectors by matching filtering the received signals at all the $K$ subcarriers, when the radar receiver has $M_r$ RF chains. Two important issues should be considered for the virtual model. (i) Increasing the number of carriers and using the virtual aperture are advantageous to improve the resolution of parameter estimation, e.g., direction-of-arrival (DOA) estimation.  (ii) More carriers will result in less power per virtual element, which in turn leads to affecting the DOA estimation performance \cite{9724206}. 

 \begin{figure}[t]
    \vspace{-1em}
     \centering
     \includegraphics[width=1\linewidth]{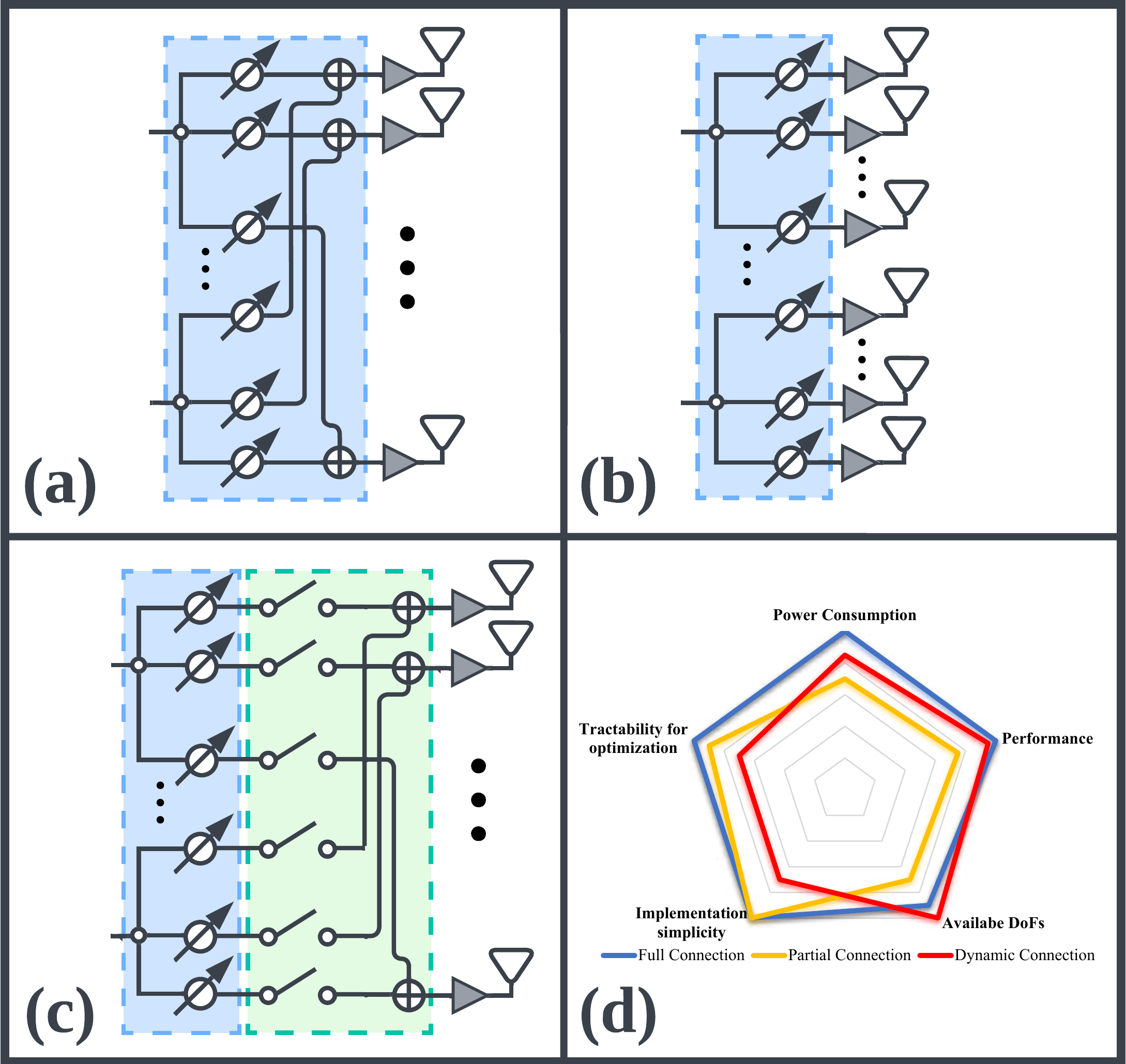}
     \caption{Comparison of the full, partial and dynamic connection analog beamformer. (a) Full Connection Structure; (b) Partial Connection Structure; (c) Dynamic Connection Structure; (d) Performance comparison of the three structure.\vspace{-1em}}
     \label{fig:HBF-System}
 \end{figure}
\subsection{Receiving Modes}
\begin{figure}[!htb]
     \centering
     \includegraphics[width=1\linewidth]{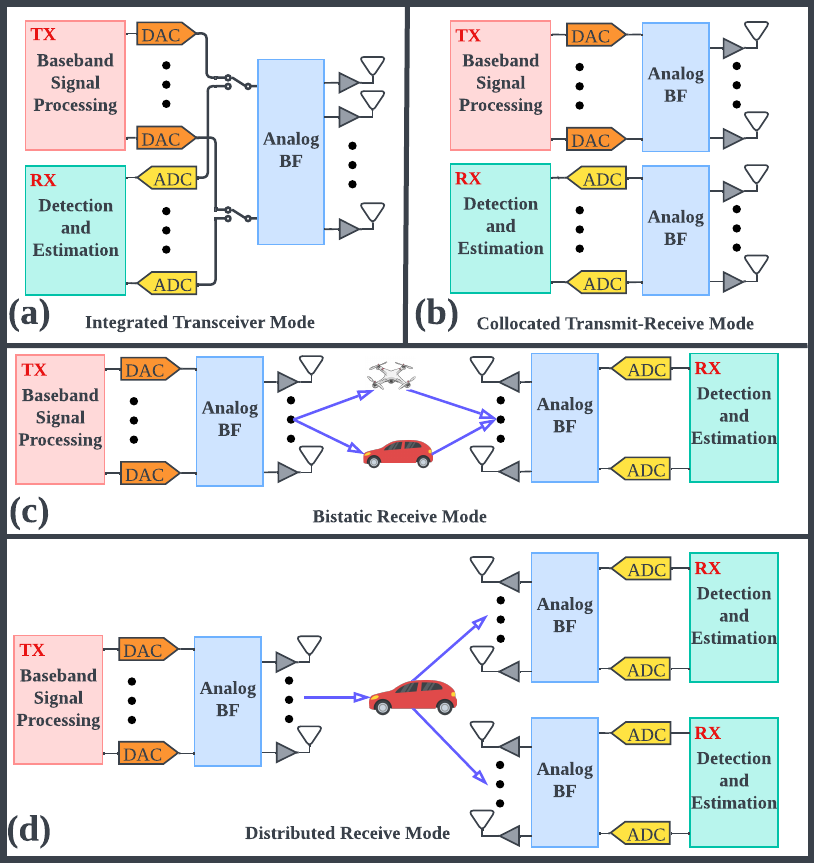}
     \caption{\textcolor{black}{Diagrams of (a) Integrated Transceiver Mode, (b) Collocated  Transmit-Receive Mode, (c) Bistatic Receive Mode and (d) Distributed Receive Mode.}\vspace{-1em}}
     \label{fig:Receiving_Model}
 \end{figure}

As shown in Fig. \ref{fig:Receiving_Model}, to support the implement of DFRC, the system can work in monostatic,  bistatic and distributed receive   modes, which are introduced as follows.

{\textbf{\color{black}Integrated    Transceiver  Mode:}} In this mode, the base station transmitter and radar receiver share the same hardware platform, as shown in Fig. 3(a), where  a   transmitter emits the DFRC waveform  to a user, and receives echoes from the surrounding  targets  simultaneously.  Since the transmitter and radar receiver share the same array, the system can only work under pulsed mode.  This mode has significant advantages of low hardware complexity and avoiding interference  caused by the leakage from transmitter, but  has shortcomings of blind spots in short range  and low communication rate. The pulse width determines the blind range.  {\color{black}In addition, since the transmitter and radar receiver share the hardware, the constraints on radar receiver should be taken into account when designing the analog precoder.} 

{\textbf{\color{black}Collocated  Transmit-Receive   Mode:}} In this mode, the transmitter and radar receiver  are closely spaced such that a target located in the far-field can
be viewed at the same spatial angle by  both. In contrast to the integrated   transceiver  mode, this mode is able to support continuous-wave signal, and has no blind spot  in  short range, but will suffer from serious interference  caused by the leakage from transmitter. Thus, it needs expensive circuitry to minimize the interference in practice. 

{\textbf{\color{black}Bistatic Receive Mode:}} In this mode, the  radar and the communication receiver share the same hardware, thereby forming a bistatic sensing mode. This  mode can support continuous-wave signal and    has a advantage  of better localization accuracy compared to the earlier monostatic case. However, this mode is  limited in the waveform to preambles and pilots.
{\color{black}Additionally, since one  array supports both radar and communication receivers, the digital combiner can be  optimized for each function, while the analog  combiner is optimized considering  both radar and communications.}  {\color{black}It is noted that this mode also includes the  uplink sensing, in which the transmitted signal is from  users, and it can be also viewed as the passive sensing \cite{9540344}.} 

{\textbf{Distributed Receive Mode:}} Herein,  the sensing signal can be received
by multiple receivers after being backscattered by the target. 
This mode can  be viewed as multiple bistatic receiving pairs. 
This model  offers spatial diversity  to   achieve  improved target detection performance and better  localization accuracy via a cooperative  processing mechanism among different receivers \cite{Blum2008Mag}. 
The HBF design for this DFRC system should consider the multi-user interference (MUI). 

\section{Design Principles of HBF in DFRC System}
 In this section,  the key performance metrics in designing the HBF for the DFRC system are discussed first, followed by the potential design approaches. Such designs are based on the formulated system performance metrics and the available communication channel state information (CSI), radar detection scenario and system requirements.
\subsection{Performance Metrics}
 In a  DRFC system,  the HBF is expected to simultaneously improve the performance  of both radar and communication functionalities  while optimizing a certain tradeoff. 
 
 {\textbf{Radar Performance Metrics:}}  Radar systems aim to detect the targets of interest,  and estimate relevant parameters, such as distance, velocity and directions. For  the diverse radar  tasks, different metrics need to be considered.  {\color{black}(1)} Particularly,  the target detection performance is quantified  by  probability of detection  ($ P_d $) under a certain probability of false alarm ($ P_{f} $). It is shown that the detection performance under Gaussian noise is positively correlated with the signal-to-interference-plus-noise ratio (SINR), which also depends on the beampattern gain. {\color{black}(2)} For multi-target detection scenario,  desired properties of the beampattern,  include multiple mainlobes, good peak sidelobe level (PSL) or integrated sidelobe level (ISL).  {\color{black}(3)}  For parameter-estimation tasks,  the Cram\'er-Rao lower
 bound (CRLB), which determines 
 the lower bound on estimation accuracy  for unbiased estimators, is  considered.  In typical radar application,   several   parameters needed to be estimated e.g.,   range,  velocity,  and direction. {\color{black}(4)}  For target characterization scenario, mutual information (MI) is also a popular metric, {\color{black}whose maximization will improve the target characterization capability 
 of a radar system}.
 
{\textbf{Communication Performance Metrics:}}  The goal of communication systems is to  enhance the rate transfer of error-free communication data over a fixed bandwidth,  this goal is commonly known as the quality-of-service (QoS). The performance metrics corresponding to the QoS includes: spectral efficiency (SE),  the mutual information\footnote{It is proved in \cite{6717211} that for a fully-digital receiver, the maximization of SE is equivalent to maximizing MI.}, symbol error rate (SER), energy efficiency (EE) and mean square
estimation error (MMSE)\footnote{Actually,  when the user adopts the MMSE receiver, the maximization of SE equals to the minimization of the MMSE \cite{5756489}.}. Apart from these metrics, for the quadrature
amplitude modulation (QAM), the   constellation range can be also selected as the figure of metric to minimize the average  SER in block HBF precoding.
\textcolor{red}{
} 
\vspace{-1em}
\subsection{Multi-Objective Optimization (MOO) Formulation}
 Typically, the problem of HBF design for the DFRC system consists of a number of objective functions, i.e., radar and communication  criteria, and is associated with a number of constraints. For a MOO problem,  the  choice of the objective function has a  significant
impact  on the trade-off performance and hence needs to selected appropriately. 

Solving an MOO problem (MOOP) implies that one should find the Pareto set of the MOOP. Numerous studies have been shown that methodologies to solve a MOOP is usually based on converting the MOO into a single-objective optimization (SOO) problem, whose optimal solution has also been proved to be in the Pareto set  of the MOOP\cite{marler2004survey}. These methodologies usually contain the 
weighted-sum, the  $\epsilon$-constraints, and
min-max  {formulations}. The detailed illustrations are provided below. 

{\textbf{Weighted-Sum {Formulation}:}} In the weighted-sum model, the multiple performance criteria are converted into a single objective by multiplying  each criterion with a given coefficient (weight). Due to the fact that different criteria may represent different performance  and have different units, the normalization of each metric must 
be undertaken prior to exploiting this  
model for solving the MOOP. Subsequently, we select the weights for different criteria to achieving the expected trade-off. It should be noted that the optimal solution of the weighted-sum problem is a Pareto-optimal one of the MOOP. The weighted-sum model is a straghtforward way to handle with the MOOP. 
%and always has a low complexity.
The  weights are non-negative and add up to 1. 
% \textcolor{red}{Different weights explore different points on the Pareto set???.}   
Nevertheless, a drawback of the weighted-sum model is that the objective is sensitive to the choice of the weights, and determination of suitable weights  to attain the satisfactory balance is difficult; 
a large number of trials with different choices of weights need to be made  to generate a satisfactory solution.
 
 {\textbf{$\epsilon$-Constraints {Formulation}:}}
 The $\epsilon$-constraints model produces  
a SOO problem, where only one 
objective function is minimized  while tackling the remaining
objectives as constraints. It is proved in \cite{marler2004survey} that the optimal solution obtained by the $\epsilon$-constraints method is  a weak Pareto optimum of the original MOOP, and under certain conditions, this solution even attain  Pareto optimality. 
 Compared to the weighted-sum method, the $\epsilon$-constraints method has a  simpler objective but more complicated constraints; this usually make the problem more challenging to solve. Similar to the choice of weights in the weighted-sum method, there is no prior  scheme corresponding to adjusting $\epsilon$. This method is normally applied to  scenarios where focus lies on only one metric  while the remaining ones are user-specified in advance.  

  {\textbf{Min-Max {Formulation}:}}  The \textit{min-max} model generates the SOO problem with $\max_i f_i({\bf x})$ being the objective function and $f_i({\bf x})$ being the $i$th objective in MOOP. A common approach to settle this problem is to introduce another variable $\eta$ to convert the  tricky  \textit{min-max}  objective into a simpler one, as $\min_{{\bf x}, \eta}~\eta, ~{\rm s.t.}~  f_i({\bf x}) \le \eta, \forall i$. However,  introducing additional constraints usually  increases   the complexity of the problem-solving. It has been shown in \cite{marler2004survey} that the \textit{min-max} method provides a necessary condition for Pareto optimal solution, but a sufficient condition only for a weak Pareto optimality. When the obtained solution is unique, it is Pareto optimal. 
\subsection{Design Approaches}
The problem of the HBF design for the DFRC system is typically nonconvex, which makes it challenging to obtain the optimal analog and digital beamformer matrices directly.
The main reasons are as follows:
%
%\textcolor{red}{For the HBF design approaches, apart from the indirect and direct designs, deep learning also used and it should be the third approach for completeness.}
%
\begin{itemize}
    \item Analog and digital precoder/combiner are coupled in the objective and constraints, which makes the resultant problem nonconvex. 
    \item Generally, taking both radar and communication metrics into consideration makes the objective nonlinear, which adds to the challenge of finding the HBF matrices. 
    \item Additionally, the constant modulus constraints on analog precoder/combiner is noconvex, which also adds to the difficulty and complexity of the problem. 
\end{itemize}

Two approaches have the potential to address these difficulties and find suboptimal solutions. \textcolor{black}{It is worth mentioning that there are some works in combining both learning and model-based strategies to form a hybrid solving approach \cite{Elbir2021JSTSP}.}
% \textcolor{red}{Before we introduce the two optimization-based approaches, it is worth pointing out that deep learning has also been explored in some works, for which please refer to \cite{elbir2019joint} and the references therein.}
% \textcolor{cyan}{
% Ref. \cite{elbir2019joint} is not related to the work of this paper. 
% Many literatures have studied the application of machine learning in HBF design (Like Geoffrey Ye Li, Feifei Gao).
% But this work is about HBF for the DFRC system.
% Machine Learning for the HBF design of DFRC has not been studied.
% }
 
{\textbf{Indirect Two-Stage Approach:}} %Since the problem of the HBF design for the DFRC system is rather complicated, 
This approach (c.f.\cite{Heath2016Overview}),  which is based on the two-stage optimization strategy,  is capable of designing the HBF in an indirect manner. In this method, the fully digital beamformer ${\bf F}^\star$ is firstly optimized by replacing the  hybrid beamformer with fully digital one in the original problem, and then the hybrid
beamformer is designed by minimizing the Euclidean distance to the obtained fully digital one. By doing so, the objective of the indirect problem is much less complicated than the original one, resulting in low complexity in computations. However, due to the indirect design, the achieved hybrid beamformer will suffer from a performance loss. Besides, it is noted that the two-stage method is not applied to solve the problem with $\epsilon$-constraints model, since the achieved hybrid beamformer that approximates the fully-digital one cannot satisfy the constraints on other metric. 

{\textbf{Direct Decoupling Approach:}} As mentioned earlier, the main difficulty in HBF design is the coupling between the analog and digital precoders/combiners. This motivates us to first decouple the hybrid beamformer to simplify the problem. For example, in \cite{cheng2021hybrid}, the objective function is a weighted summation of the multi-user MMSE  and
radar spatial spectrum matching error (SSME). To decouple the analog and digital beamformers, a set of auxiliary variables ${\bf Y}_{k,u}={\bf F}_{\rm RF}{\bf F}_{k}$ are introduced, such that the complex objective function can be decomposed into $UK$ sub-functions, and each sub-function depends only upon a variable ${\bf Y}_{k,u}$. This enable us to iteratively optimize $( \{{\bf Y}_{k,u}\}, {\bf F}_{\rm RF}, \{{\bf F}_{k}\})$  under the consensus alternating direction method of multipliers (consensus-ADMM) framework\cite{boyd2011distributed}. %In this framework, variables $\{{\bf Y}_{k,u}\}$ are updated in a  parallel manner. Subsequently, ${\bf F}_{\rm RF}$ with constant modulus constraint is optimized with the aid of the Riemannian Conjugate Gradient (RCG) algorithm. With   $( \{{\bf Y}_{k,u}\}, {\bf F}_{\rm RF}  )$ known, closed-form solutions of  $\{{\bf F}_{k}\} $ can then be attained in parallel. 
\section{Performance analysis }
In this section,  we illustrate the performance of the DFRC system with different HBF architecture and design approaches through numerical simulations.

\begin{figure}[t]
     \centering
     \includegraphics[width=1\linewidth]{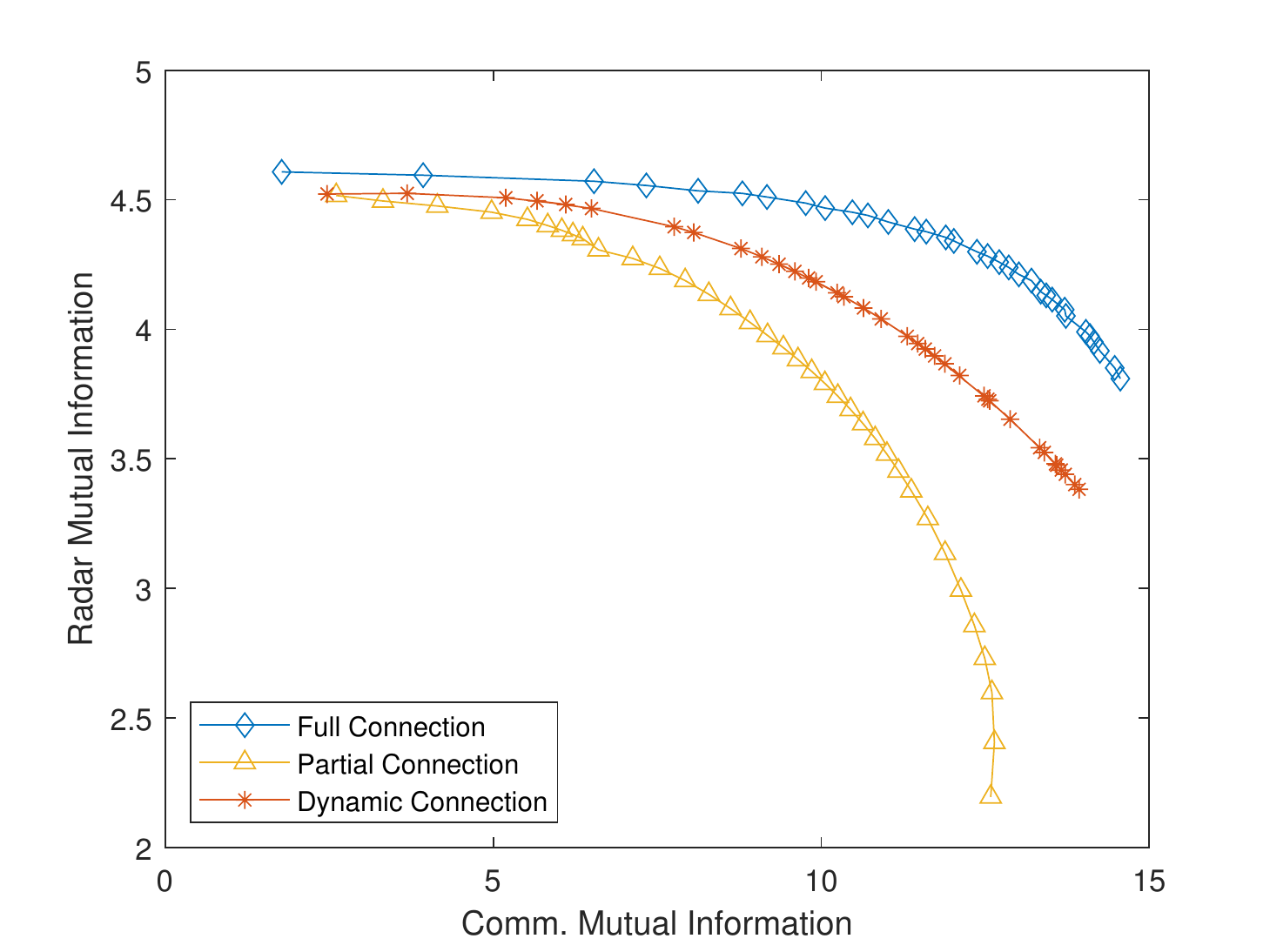}
     \caption{The trader-off between radar mutual information and communication information with different PSs network structure.$N_{t} = 32$, $N_{RF}^t = N_{s} = 4$.}
     \label{fig:Sim_parato}
 \end{figure}

We first demonstrate the  trade-off of the DFRC system between the radar and communication functionalities as shown in Fig. \ref{fig:Sim_parato}. In the considered scenario, the optimization problem is formulated in the weighted-sum form. By tuning the weight parameters with a small stepsize, we obtain the Pareto bounds.
It is observed that the radar mutual information decreases monotonically with an increase in the communication mutual information under all the structures, which reveals the  Parato optmaility inside the DFRC system on the radar and communication aspects. This Parato optimality further implies that to achieve an appropriate trade-off between sensing and communication, selection of the weight parameters is crucial . Moreover, these results also indicate that an appropriate choice of  the HBF structure can trade-off between the hardware complexity and DFRC performance.

\begin{figure}[t]
     \centering
     \includegraphics[width=1\linewidth]{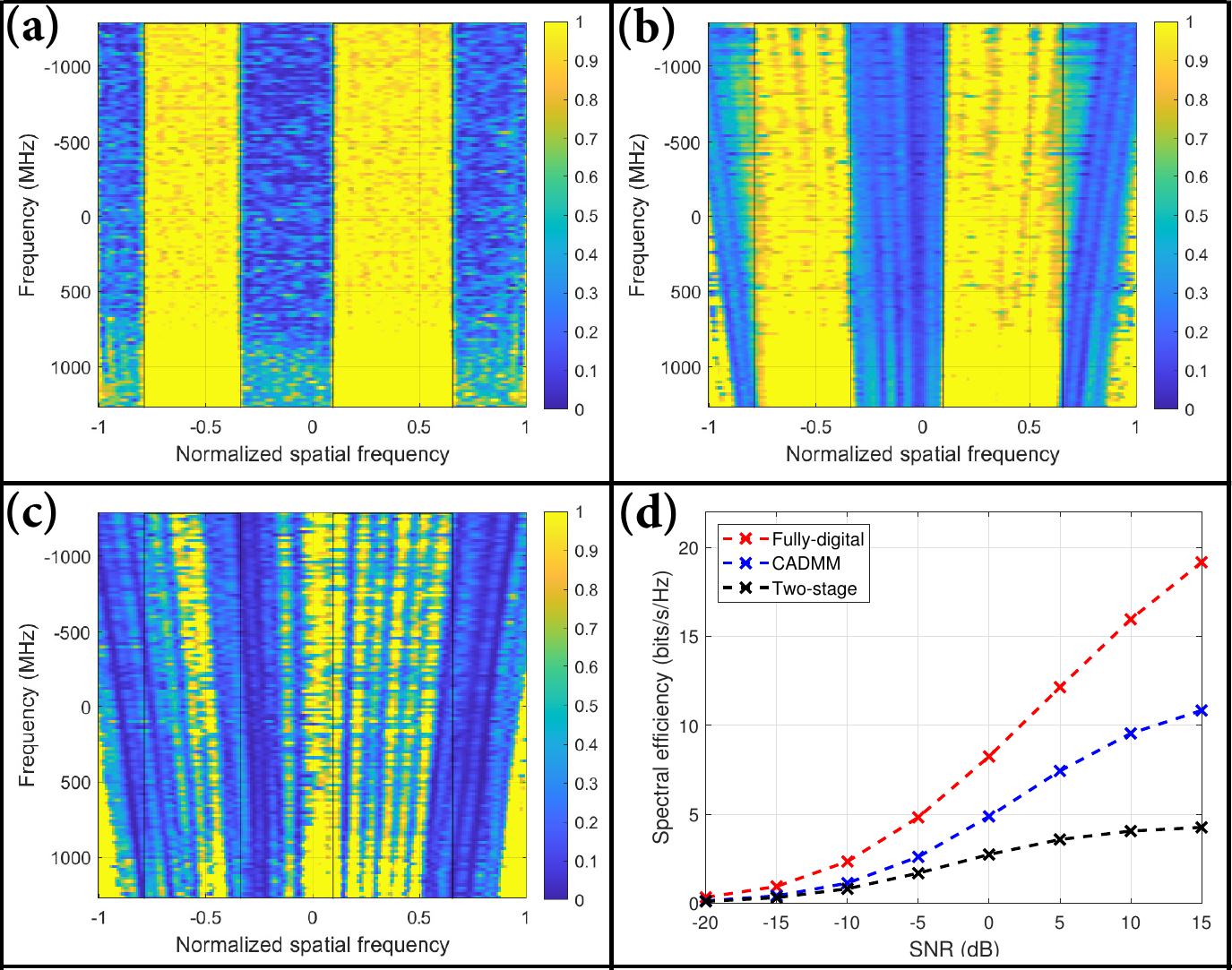}
     \caption{The DFRC system is assumed to be equipped with $N_{{t}}=32$ antennas and $N_{\text{RF}}=4$ RF chain serving $U=4$ downlink users. The main-lobe region is ${{\bf{\Omega }}_{main}} = [ - 0.7891, - 0.337] \cup [0.0939,0.657]$. (a) The space-frequency spectra for Fully Digital case; (b) The space-frequency spectra for CADMM case; (c) The space-frequency spectra for Two Stage case; (d) The spectral efficiency versus the SNR for different methods.}
     \label{fig:Sim}
 \end{figure}

Next, we provide simulation results of different hybrid beamforming design schemes as shown in Fig.\ref{fig:Sim}.
In this simulation, we consider the wideband OFDM based DFRC system serving multiple downlink users and detecting radar targets, where the simulation setting and more details can be found in \cite{cheng2021hybrid}.
For the comparison purpose, the DFRC system with the fully digital structure is included as the performance upper bound.
It is clear to see that the two-stage method suffers from severe performance loss in terms of both radar beampattern and communication spectral efficiency. In contrast, with the direct design, hybrid beamforming for the DFRC system achieve better performance consistenly at all SNR levels.

 \section{{New Frontiers in Advanced HBF DFRC Systems}}
% \textcolor{red}{Red notes are the revisions or comments. In a summary: (1) Please double check some technical details or terminologies to ensure correctness and rigorousness (2) For each topic, please refine the logic and language (3) Please emphasize either the research gain or challenges which are somehow unique in HBF DFRC. This is important to distinguish our works from the counterparts in pure communications!}
 
%Despite the growing interest on the HBF for DFRC, the relevant design  still has numerous  challenges for future researches on the HBF-based DFRC, as discussed below.
%
{From a practical perspective, HBF is a promising technology to be implemented in the upcoming DFRC system. However, to achieve both the sensing and communication functionalities, the vanilla HBF might not fulfill the requirements and achieve optimal tradeoffs. Hence, advance HBF techniques should be developed on various aspects such as architecture, signal processing, and technology convergence; these will be discussed in this section.}
\subsection{HBF in Wideband  DFRC Systems}
 In order to maintain reliable robustness
against multipath fading for communication and to improve 
estimation accuracy for sensing, wideband is necessary. Specifically, 
wideband  OFDM system  can be regarded as a special case of multi-carrier system, when the number of carriers is very large, the system presents the  wideband property. Different from the narrowband model, the delay of wideband system is related to the frequency, this results in \textit{frequency-dependent} beampattern \cite{cheng2021hybrid}. Moreover, for the analog beamformer, different frequencies need different phase shifts. As a result, the standard HBF architecture based on  PSs cannot deal with the beam-squint problem   caused
by the large number of antennas and bandwidth, as well as the small number of RF chains. The beam-squint  problem will lead to the  beam pointing offset and beam split effect, in which different subcarriers have different  spatial directions.   The  beam-squint  problem will result in a situation where  only the beams around the expected subcarrier are able to attain a high gain, while  the beams at other subcarriers suffer from  a serious loss in gain, which has not been well investigated in wideband HBF-based DFRC systems  \cite{Elbir2021JSTSP}. 
In order to  mitigate this effect, combining the time delay network  with analog PSs is able to be introduced to general \textit{frequency-dependent-like} PSs to form the \textit{frequency-dependent} beams. Therefore, research on the  delay-phase hybrid beamforming
design and corresponding performance analysis  should be carried on in future work for 
wideband DFRC systems.

\subsection{HBF in DFRC Systems With Low-Resolution ADCs and DACs}
In mmWave systems, the high sampling
rate and high quantization bit of ADCs and DACs will result in high power consumption and cost (e.g., a high resolution ADC $\ge$ 8 bits consumes several Watts \cite{761034}). The HBF structure is capable of reducing the power consumption and cost via adopting a limited of RF chains. Besides the HBF structure, the utilization of low-resolution ADCs/DACs is another ingredient for mmWave system. Although the low-resolution ADCs/DACs have advantages of saving the cost and  power consumption, they may lead to severe performance loss. 
In communication applications, with low-resolution ADCs, the performance of HBF and fully digital beamforming has been compared. The results show that at low SNR regions, the fully digital beamforming with low resolution ADCs achieves the same performance to the HBF with ideal ADCs. In addition,  the HBF with coarse quantization attains better performance-EE trade-off.  However, there is a lack of research works on joint optimization of HBF and ADCs/DACs resolutions for mmWave DFRC systems. 
%Clearly, it is looking forward to investigate the combination of the two efficient approaches for the DFRC systems. 
Further, the analysis of low-resolution ADCs/DACs in DFRC will help us come to a new understanding related to the HBF design issues, but which is almost blank currently. For example, in a specific DFRC application, it is not trivial to determine the appropriate RF chain number, bit-resolution of ADCs/DACs and  PS resolutions to achieve an excellent trade-off among radar performance, communication performance and hardware efficiency. 

\subsection{RIS-Assisted  HBF DFRC Systems}
Reconfigurable Intelligent Surface (RIS) consists of multiple passive elements, each of which can introduce phase shift independently and hence alter the radio propagation environment. From a signal processing perspective, it is essentially shaping a passive beamforming on the impinging signals. Therefore, it is appealing to integrate RIS into the DFRC system to enhance system performance including beam accuracy, coverage  and spatial diversity, etc. The resulting problems are usually formulated as a joint design of both the active (i.e., HBF on the station) and passive (i.e., RIS) beamforming, which requires efficient optimization algorithms especially in the context of large-scale DFRC deployment. Furthermore, RIS has been evolving from the standard passive, discrete and narrowband case to an advanced one which exhibits active, holographic and wideband properties. Correspondingly, the joint design of RIS and HBF becomes even more challenging, thereby warranting an investigation in the HBF DFRC settings. For example, the beam-squint effect overcome by wideband HBF shall re-appear, unfortunately, after reflecting by a narrowband RIS. Apart from the design aspects, it is meaningful from an analytical perspective to pursue the goal of proving the theoretical bound on performance gain in such a system $-$ a matter less investigated in literature compared to the design aspects.
\subsection{OTFS Modulated HBF DFRC Systems}
In high-mobility scenarios, the conventional OFDM suffers from severe Doppler spread and thereby serious inter-carrier interference (ICI). Recently, a novel two-dimensional modulation, i.e., orthogonal time frequency space (OTFS), has been regarded as a promising alternative to the OFDM \cite{9508932}. Different from the OFDM in the time-frequency domain, the OTFS modulation works in the delay-Doppler (DD) domain. Thus, when considering the OTFS-based DFRC systems, the HBF design may face the challenges. For example, how to achieve appropriate radar and communication objective functions based on OTFS modulation signaling with the HBF structure?  Moreover,  for the HBF in OTFS-based DFRC systems, the digital beamforming is designed  for each DD block, while the analog beamforming needs to be  optimized for the whole bandwidth. The corresponding  problem  has a stronger  coupling between the analog and digital beamformers.  Additionally, since the number of DD blocks are very large, it is of utmost necessity to  seek a low-complexity solution.  %The consensus-ADMM method will be a promising candidate.  
In summary,  it is of great interest to examine performance of the  OTFS-based DFRC systems with the HBF structure for future works.
\section{Conclusion}
Hybrid beamforming aided DFRC has the potential to significantly improve the energy/spectral efficiency and enable ubiquitous connectivity in future B5G and 6G networks. In order to imperatively understand and tackle the challenges associated with it, this article presented a comprehensive overview of DFRC with hybrid beamforming technology.
Starting from the fundamentals of the hybrid beamforming, we first introduced the system model of the DFRC with hybrid beamforming, including the system structure and DFRC transceiver architecture.
Then the design principles for hybrid beamforming  incorporating DFRC system was introduced based on performance metrics and prevalant approaches.
% Finally, we clarified the challenges and discuss the open issues as future research directions.
We then present representative results to highlight the princinples  and finally outline several future research directions that can facilitate the full potential of DFRC with hybrid beamforming technology.

\appendices

\ifCLASSOPTIONcaptionsoff
  \newpage
\fi



%\bibliographystyle{IEEEtran}

\footnotesize
\balance
 \bibliographystyle{IEEEtran}
 \bibliography{IEEEabrv,stan_ref}

\end{document}